\newif\ifshowedit
\newif\ifisdraft

\showeditfalse
\isdraftfalse

\ifisdraft

	\documentclass[journal,11pt,onecolumn,draftclsnofoot,]{IEEEtran}
\else
	\documentclass[10pt,final]{IEEEtran}
\fi

\usepackage{graphicx}
\usepackage{dcolumn}
\usepackage{bm}
\usepackage{mathrsfs}
\usepackage{float}
\usepackage{amssymb}
\usepackage[cmex10]{amsmath}
\usepackage{amsthm}
\usepackage{epstopdf}
\usepackage{cite}
\usepackage{xr}
\usepackage{color}
\usepackage{soul}

 \DeclareMathOperator{\sgn}{sgn}

\ifshowedit
	 \newcommand{\add}[1]{{\color{blue}#1}}
 	\newcommand{\del}[1]{{\color{red} \setstcolor{red}\st{#1}}}
 	\newcommand{\deleq}[1]{{\color{red} \setstcolor{red}#1}}
\else
	\newcommand{\add}[1]{#1}
	\newcommand{\del}[1]{}
	\newcommand{\deleq}[1]{}
 \fi
 
\begin{document}


\title{A Biased Look at Phase Locking:\\Brief Critical Review and Proposed Remedy} 
%

\author{\IEEEauthorblockN{Christopher K. Kovach\IEEEauthorrefmark{1},
\IEEEauthorblockA{\IEEEauthorrefmark{1}Department of Neurosurgery,
University of Iowa Hospitals and Clinics, Iowa City, IA}
\thanks{ Corresponding author: C.K. Kovach }}}


\date{\today}


\IEEEtitleabstractindextext{%
\begin{abstract}
A number of popular measures of dependence between pairs of band-limited signals rely on analytic phase. 
A common misconception is that the dependence revealed by these measures must be specific to the spectral range of the filtered input signals. 
Implicitly or explicitly, obtaining analytic phase involves normalizing the signal by its own envelope, which is a nonlinear operation that introduces broad spectral leakage.
We review how this generates bias and complicates the interpretation of commonly used measures of phase locking. 
A specific example of this effect may create spurious phase locking as a consequence of nonzero circular mean in the phase of input signals, which can be viewed as spectral leakage to 0 Hz.
\del{We describe an iterative recentering procedure to correct this problem is described, but it is also noted that recentering fails}
\add{Corrections for this problem which recenter or uniformize the distribution of phase may fail} when the amplitudes of the compared signals are correlated.
To address the more general problem of spectral bias, a novel measure of phase locking is proposed, the amplitude-weighted phase locking value (awPLV). 
This measure is closely related to coherence, but it removes ambiguities of interpretation that detract from the latter. 

\end{abstract}
}

\maketitle

\IEEEdisplaynontitleabstractindextext
\IEEEpeerreviewmaketitle



\section{\label{sec:level1}Introduction}

\IEEEPARstart{M}{easures} of phase locking are rapidly becoming a standard tool in the analysis of biophysical signals, especially so in applications to neurophysiology, where they are used to study interactions between anatomically separated neural populations \cite{nunez1997eeg, lachaux1999measuring,mormann2000mean} as well as cross-frequency coupling within responses from the same population \cite{canolty2010functional, tort2010measuring}. 
Their burst in popularity has been followed by a more gradually dawning awareness of the many pitfalls that accompany their use \cite{zaveri2000effect, schiff2005dangerous, guevara2005phase, chavez2006towards,celka2007, kramer2008sharp, aydore2013, aru2015untangling, srinath2014effect, bastos2015tutorial, lowet2016quantifying}. 
For example, recent entries to the body of cautionary literature \cite{kralemann2008phase,Cox14,van2015phase} point out how non-uniformity of the distribution of phase in the input signals may create spurious results in many of the most commonly used indices of phase locking.
\add{This problem may be identified as one of a more general family of biases related to the spectral interpretation of phase.}  
\del{Although the authors of this work, van Driel et al, describe the problem in applications to cross-frequency coupling, their point applies generally as well.} 

\del{Here we consider the issue from the spectral domain, where the bias described by van Driel et al. can be understood as a specific instance of a more general class of spectral leakage biases, all of which}\add{Viewed from the spectral domain, it becomes clear that all members of this family} share the same culprit:  the transformation through which phase is obtained, which divides an analytic signal by its own envelope. 
The nonlinearity of this transformation induces spectral leakage across a broad range of frequencies\add{, and the resulting unit-amplitude representation of phase is in general not a bandlimited  function \cite{Papp71,picinbono1997on}.}  
Moreover, consequences of this are not \del{limited}\add{confined} to measures that explicitly invoke envelope normalization, but also extend to those that \del{directly} extract phase from the argument of the analytic signal, for which one may consider normalization as implicit.  
The measures in question include phase-locking values (PLV), also known as phase coherence \cite{lachaux1999measuring}, phase-amplitude coupling \cite{canolty2010functional} and information-theoretic measures derived from the empirical distribution of phase angles \cite{tort2010measuring}, along with others \add{that correct for small-sample bias} \cite{vinck2010pairwise}. 

One class of measure\del{, not considered by van Driel et al. in their analysis of the problem,} is immune to spectral biases of this type: those obtained directly from signal cross spectra, of which coherence is the prime example. 
By avoiding amplitude normalization, coherence circumvents the underlying spectral distortions. 
However, the ambiguity of \del{coherence}\add{its interpretation}, which depends on both phase locking and amplitude correlation, has dampened its popularity\del{and helped to }\add{, motivating} \del{the wide adoption of }\add{the aforementioned} alternatives that discard amplitude \cite{lachaux1999measuring}.

After reviewing the origin of spectral leakage in analytic phase, \del{we consider an}\add{another} alternative \del{measure}\add{is considered}, the amplitude-weighted phase locking value (awPLV), which combines some of the favorable qualities of coherence with those of PLV. 
Unlike coherence, awPLV always yields a value of 1 for perfectly phase-synchronized signals \add{(meaning here two signals whose phase difference remains constant over time)}, regardless of amplitude correlations. It improves on PLV and other phase-only measures by avoiding the spectral distortion responsible for the aforementioned biases. 
It also \del{overcomes}\add{addresses} another \add{potential} drawback of PLV, which is that PLV weights all samples equally, including those in which the signal amplitude approaches zero, for which phase is either poorly defined or likely to be dominated by background noise. 
Like coherence, awPLV weights the phase of each sample according to the product of the input signal amplitudes; its numerator therefore contains a cross-spectral estimate. 
It \del{differs}\add{diverges} from coherence \del{in}\add{by} normalizing \del{by}\add{with} the sum of amplitude products rather than the product of separate root-mean-square amplitudes. 
Weighting this way allows the resulting measure to be viewed, like PLV, as an average of unit-length phase vectors, but a weighted average.

\subsection{Nonlinearity and Phase}
\label{section:nonlineartyAndPhase}
Throughout the following discussion, spectral broadening related to amplitude normalization will be described as ``spectral leakage,'' and its effect as a form of ``bias.'' 
Some readers may rightly object to this terminology: spectral leakage bias normally describes a set of artifacts extrinsic to the signal, related to multiplying the signal with \del{a predetermined}\add{an externally determined} window function, not, as with amplitude normalization, a ``window'' pulled from the signal itself.
Our use (or rather abuse) of this terminology calls for more justification.
It should be emphasized at the outset that spectral broadening from amplitude normalization is neither an artifact nor inherently a flaw in conventional measures of phase.
\add{Though} phase synchronization \add{may arise through linear interactions, ``pure'' phase-phase interactions, which are not affected by signal amplitude, are nonlinear}, so it would be a mistake to assume that the spectral constraints of linear interactions must apply generally to phase locking. 
\add{In many settings, this nonlinearity is what makes phase interesting \cite{Freeman20042077}.
On the other hand, it is very difficult to resist the urge to think of phase as a spectrally localized quantity, given that the underlying signals are often band-limited and the rate of change of phase defines the instantaneous frequency.}

\del{This is self-evidently the case for for so-called \it n:m \rm phase locking, when phase in one signal tends towards a stable ratio with that of another, each of which may occupy entirely different spectral ranges 
Nevertheless, most commonly used measures of phase locking address only the simple 1:1 case, while the more general case is often handled by combining the same measures with specific transformations of the input signals.
Considerations in the simple case will therefore usually extend more generally.
We will likewise only address the simple case applied to signals that are assumed to be band-limited within an overlapping spectral range.}

\del{Even in this simple case, however, spectra after amplitude normalization are not in general band-limited.
Phase interactions are therefore not inherently narrowband, even when the underlying signals are. 
On one hand, the nonlinearity implied by this fact is what makes phase interesting in some contexts 
On the other, it is very difficult to resist the urge to think of phase as a spectrally localized quantity, given that the underlying signals are usually band limited.
The time derivative of phase also gives the ``instantaneous frequency,'' whose very name implies  spectrotemporal localization.
But it is important to draw the distinction between the instantaneous properties of an analytic signal and its spectrum 
; this distinction is essential to the ``bias'' we  consider here.}

A detailed consideration of the spectral interpretation of instantaneous phase \add{is} not the \add{primary} goal of the present work, however. 
\add{A large body of work already addresses this topic \cite{Papp71, Mandel74, boashash1992estimating, Loughlin97,Loughlin1997b}, which is also closely related to the longstanding question of the relationship between instantaneous frequency and and the signal spectrum}\cite{ville1948theorie,Cohen1989,picinbono1997on}. 
The present aim is instead to describe conditions under which phase locking might be safely viewed as a measure of frequency-specific dependence between pairs of signals, in the same way as measures of linear dependence derived from the cross-spectrum. 
For this purpose, spectral broadening in analytic phase will be treated as a form of spectral leakage bias, although readers who prefer more nuanced terminology might wish to replace ``bias'' with ``misinterpretation of the spectral nature of phase locking.''

\add{\subsection{Organization}
\label{section:organization}
Sections \ref{PhaseMeasures} and \ref{SpectBias} are intended as a tutorial introduction. Section \ref{section:decbias} presents the basic problem of bias related to non-uniform phase distributions, while section \ref{section:ampmod} explains how and why previously described solutions to this problem may fail in the presence of correlated amplitude modulation.  Section \ref{Remedies} considers some methods for overcoming the problem, with emphasis on the awPLV.}


\section{Measures of Dependence in Complex Signals}

\subsection{Complex and Analytic Signal Representations}
\label{ComplexAnalytic}
Most commonly used measures of phase locking are derived from complex signal representations of the type obtained with a combination of band-limiting filter and Hilbert transform or \del{formally equivalent}\add{related} procedures such as a continuous wavelet transform \cite{le2001comparison, bruns2004fourier}, short-time Fourier transform \cite{gabor1946theory} or complex demodulation \cite{bingham1967modern, allen1977unified,kovach2015demodulated}. 
In all cases, the resulting \del{analytic}\add{complex} signal can be understood as the outcome of a band-limiting filter applied to the original real-valued signal.\del{, which suppresses negative frequencies.} 

\add{It is common to think of the result as analytic, meaning that it contains no negative frequencies, but this restriction is not essential.
The important feature is rather that the filter be of a bandpass type, attenuating energy outside a single spectral band. 
Because its spectrum must be symmetric about 0 Hz, a real-valued signal fulfills this condition only if it is lowpass filtered; for this reason, the result is in general complex-valued, even when not strictly analytic.
In the following discussion, the term ``analytic signal'' will be used loosely to refer to any complex-valued signal obtained with a \add{single-band} filter.
The conventional procedure combines an ordinary bandpass filter, $g$, with a Hilbert-transform-approximating quadrature filter, which suppresses energy in the negative band, so that the complex-valued result is the filtered signal in the real part and its (approximate) Hilbert transform in the imaginary part:
\begin{eqnarray}
X(t)=g*x(t)+i\mathscr{H}\{g*x(t)\}
\end{eqnarray}
} 
\del{Because this transformation} \add{In all cases, the result} may be understood as \del{an application of}\add{the outcome of} a simple band-limiting filter, \add{which will therefore} \del{it is linear, and linearity implies that the result will}contain no spectral component lacking in the original signal.  

\add{The same nonlinearity of phase that complicates its spectral interpretation also makes its meaning unclear when the signal is the sum of many components.
One purpose of bandpass filtering in this setting is to isolate components in the hope of obtaining an interpretable result. 
While bandpass filtering is the simplest and most common approach the problem, it is by no means the only one \cite{huang1998empirical,mallat1993matching}.
To avoid questions tangential to the present aim it will therefore be assumed that the signal of interest, $x$, falls within the passband of $g$, so that $g*x(t)\approx x(t)$, except where otherwise specified.
Finally, it is very often useful to represent the analytic signal in a form that makes phase explicit, as the product of a positive real-valued envelope function $A(t)$ and a complex unit-magnitude valued phase: }
\begin{eqnarray}
	X(t)=A_x(t)e^{i\phi_x(t)}
\end{eqnarray}
with $A_x=|X(t)|$ and $\phi_x = \mathop{\mathrm{arg}}(X)$.

\del{Phase locking analyses are frequently carried out after passing the signal through multiple band-limiting filters with varying center frequencies and/or bandwidths in order to  characterize phase dependence in different spectral ranges. 
So-called time-frequency decompositions add center frequency as an index into different filter bands, $X(\omega,t)$, where 
each component indexed by $\omega$ is a separate band-limited analytic signal.
Such representations are the starting point of most stationary and time-varying spectral and cross-spectral estimators. }

\del{Although it is perhaps more intuitive to think of such time-frequency decompositions ``vertically,'' as spectral decompositions applied to a sliding local time window, it is often useful and sometimes conceptually simpler to think of them ``horizontally'' in this way, as a series of analytic signals produced by band-limiting filters. 
A remarkable fact is that both ways of thinking about the problem give fundamentally equivalent signal representations 
Nevertheless, it is often simpler to view the problem of spectral estimation from the filter-bank perspective because basic spectral quantities, like cross-spectral estimates, are computed separately within bands. In this perspective, to understand properties of a spectral estimator at all frequencies, it suffices to consider operations within a single band 
The main advantage of this formulation is that that it appeals directly and explicitly to the close relationship between spectral estimation and operations on bandpass filtered analytic signal decompositions, unifying what are often treated as distinct groups of methods. }

\subsection{Cross Spectra and Coherency}
\label{section:cspect}
\add{Instantaneous} linear relationships between \del{analytic}\add{complex} signals are quantified in the same way as for real-valued signals through \del{the complex-valued analog of covariance}\add{measures derived from the second moment at zero lag:} 
\begin{eqnarray}
\label{EQ_cspec}
	\add{R_{XY}(t) = \mathrm{E}\left[ X(t)Y^*(t) \right]}
\end{eqnarray}
When $X$ and $Y$ are stationary,  this expectation assumes a constant value, which may then be estimated by integrating over time:
\begin{eqnarray}
\label{EQ_cspec_est}
	\hat{S}_{XY} =  \add{\frac{1}{T}}\int_T{
				X(t)Y^*(t)
				\mathop{dt} 
			}
		     = \add{\frac{1}{T}}\int_T{
				A_xA_ye^{i(\phi_x-\phi_y)}
				\mathop{dt} 
			} 
\end{eqnarray}
\del{For the time-frequency example considered above, this}
\add{Because the analysis filter in the present case, $g$ introduces timing uncertainty, the relationship revealed by this is not properly ``instantaneous,''  rather its timing is ambiguous according to the uncertainty relationship. 
Eq. (\ref{EQ_cspec}) instead may} be regarded as an estimate of a cross-spectr\del{um}\add{al coefficient} within \del{each band}\add{the band of the analysis filter, $g$\cite{koopmans1995spectral}. } 
\del{It is virtually always the case that the analytic signals in question are band-limited, for which reason} 

\del{It is virtually always the case that the analytic signals in question are band-limited, for which reason there is no need to worry about subtracting the signal averages, as one would in computing sample covariance for a real-valued signal. 
Mean centering is redundant with the bandpass nature of the analytic signal, which normally excludes any DC (i.e.  0 Hz) component from the original signal. 
A caveat should be added for non-stationary signals when averaging over selected time points, as when computing coherence across trials; in such cases there is a clear reason to subtract the corresponding means, and no reason to recommend against it. For the sake of avoiding tangential issues, however, we will assume we are dealing with ``wide sense'' stationary signals, whose spectra remain constant over time, including with respect to the mean, unless otherwise specified.
}

Measures \add{of dependence} derived from $\hat{S}_{xy}$ in essence reflect the spectral overlap between signals within the originating band. 
This fact follows from Plancherel's theorem:
\begin{eqnarray}
\label{EQ_Planch}
	 \int_{-\infty}^\infty{
	 	  X(t)Y^*(t) \mathop{dt}
		  }  
	  = \int_A^B{
	 	  \tilde{X}(\omega)\tilde{Y}^*(\omega) \mathop{d\omega}
		  }			
\end{eqnarray}
where the support of $\tilde{X}(\omega)\tilde{Y^*}(\omega)$ is restricted to the band $[A,B]$. 
Two signals with non-overlapping spectra will therefore always have a cross-spectrum of 0. 
Signals whose spectra overlap but are otherwise uncorrelated will yield cross-spectral estimates subject to small-sample bias tending to 0 with increasing sample size. 

Just as one quantifies the degree of instantaneous dependence between real-valued sequences with Pearson's correlation coefficient, a measure of dependence for analytic signals is obtained by scaling the cross-spectral estimate, \emph{coherency}:
\begin{eqnarray}
\label{EQ_coh}
	C_{xy}=\frac{
		 	S_{xy}  
		 }{
		  \sqrt{ S_{xx}S_{yy} }
		}		
\end{eqnarray}
Coherency differs from its real-valued analog in one critical respect: its nature is composite, blending two distinct aspects of the dependence between analytic signals. 
The first relates to the correlation of envelopes, and the second to the consistency of phase difference between the signals, $\Delta\phi_{xy}=\phi_x(t)-\phi_y(t)$, that is, their degree of phase locking. 
The overall degree of dependence may be quantified by the magnitude of the coherency, $|C_{xy}|$ (referred to as coherenc\emph{e}) while the characteristic difference of phase between signals is given by the argument (phase angle), $\theta_{xy}=\mathop{\mathrm{arg}}C_{xy}$. 

Because of its composite nature, the meaning of coherence is inherently ambiguous.  A given value might reflect either correlated amplitudes or phase locking. 
Moreover, the contribution of each term depends on an interaction with the other. 
When the signals remain in a perfectly \del{consistent}\add{constant} phase relationship \add{($\Delta\phi_{xy}(t)=c$)}, coherence reduces to an uncentered correlation of the amplitudes. 
In such cases, it will tend to vary around some intermediate value when amplitudes are uncorrelated, but \add{it} can assume any other value depending on the variance and \del{skew}\add{skewness} of the amplitude distributions. 
On the other hand, two signals that have no phase relationship will give coherence that tends to 0, regardless of amplitude correlation. 
This ambiguity represents a significant drawback to an otherwise appealingly simple measure.

\subsection{Measures on Phase Only}
\label{PhaseMeasures}

The problems with coherence motivate alternative measures that are more easily interpreted. 
A seemingly straightforward way to address the issue is to examine dependence between the phases of analytic signals in isolation from amplitude. 
Measures of pure phase dependence discard information about signal amplitudes by dividing each signal by its own modulus, thus normalizing away amplitude fluctuations:
\begin{eqnarray}
\label{EQ_Phi}
	 \Phi_x(t)= \frac{ X(t) }{|X(t)|} 
			  = e^{i\phi_x(t)}	
\end{eqnarray}
The simplest of these pure phase measures, phase-locking value (PLV), or phase coherency, computes coherency as in (\ref{EQ_coh}) from the amplitude normalized signals, giving the expectation of the phase difference vector, $E\left(e^{i\Delta\phi_{xy}}\right)$ \cite{ lachaux1999measuring}.
The magnitude of this vector remains at 1 when the phase difference between the signals is perfectly stable and tends to zero when there is no consistent phase relationship. 
This measure also has an  intuitive geometric interpretation as the degree to which a sample of unit length phase vectors, representing the \add{angular} difference, align in the same direction.

One drawback of PLV is that it addresses \del{only linear relationships}\add{only second-order dependence} between the phase vectors. 
More general forms of dependence can be revealed through information-theoretic metrics applied to the \add{bivariate} distribution of phase angle\add{s}\del{ differences}, such as Kullback--Leibler divergence or the Kolmogorov--Smirnov test. 
However, both PLV and these alternatives are fundamentally measures on the same distribution of phase angles, meaning that differences between them reflect the nature of the measure rather than any property of the quantity measured. 
For this reason, one may view amplitude normalization as implicit in measures based purely on phase,\del{as are its} \add{along with any} spectral consequence.

\add{
\subsection{Nonstationary Extensions}
\label{section:nonstationaryApp}
Discussion has so far been confined to stationary and ergodic signals, but it is often of interest to consider applications in which stationarity is not assumed. 
One simple approach to the nonstationary case accounts for time dependence by introducing time-weighting to the estimator. 
Formulating Eq. (\ref{EQ_cspec_est}) as a Riemann-Stieltjes integral over the relevant weighting measure allows the extension to apply both to continuous and atomic weighting.
Such weighting may, for example, describe a subsampling of the signal used  to reveal phase-locking at specific delays with respect to some externally defined series of events, allowing for event-related phase-locking analyses. 
This point will be considered again briefly in section \ref{section:genweight}.
} 

\begin{figure*} 
\includegraphics{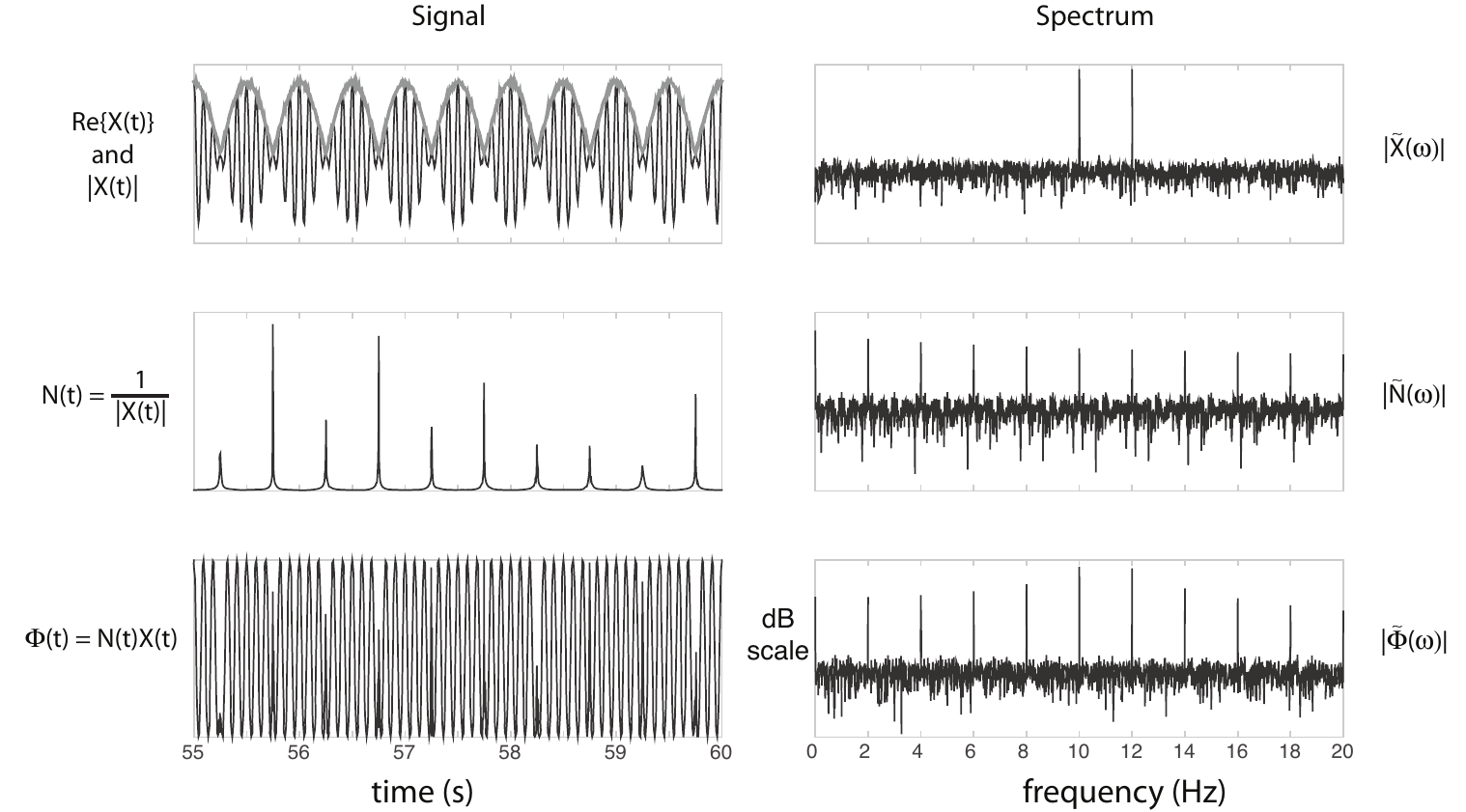}
\caption{\label{fig:beat_examp} 
Spectral consequence of amplitude normalization.  
\emph{Top panels}: the test signal, $X(t)$ is composed of two sinusoids at 10 Hz and 12 Hz, whose summation creates an amplitude modulated signal with center frequency 11 and modulation frequency of 2. 
\emph{Middle panels}: the inverse amplitude function, $N_x(t)$, exhibits spikes where the amplitude envelope approaches 0, for which reason its spectrum, $\tilde{N}(\omega)$, includes harmonics of the modulation frequency.   
\emph{Bottom panels}: As a result of the convolution property of the Fourier transform, multiplying $X$ by its inverse modulus to obtain $\Phi$ gives a signal that includes harmonics and subharmonics of the test signal spaced at 2 Hz. These features create the potential for spurious phase locking if they overlap with features of spectrally unrelated signals.
}
\end{figure*}

\section{Spectral Bias from Amplitude Normalization}
\label{SpectBias}
We have just seen that measures derived from cross-spectral estimates represent the dependence between signals within their originating bands. 
The same is not true for measures based purely on phase. 
This fact follows from the nonlinearity of amplitude normalization, which may induce spectral components outside the original signal bandwidth. 
Two signals that initially shared no common spectral features, with coherence strictly zero, may therefore exhibit nonzero coherence after the normalization (i.e. phase locking) should any of the induced spectral features happen to overlap. Next, we consider how this comes about.

\subsection{Origin in the Fourier Domain}

Normalization entails multiplying the original band-limited analytic signal, $X(t)$, by a function obtained from the inverse of its \del{amplitude}\add{norm, $N_x(t) = |X(t)|^{-1}$}. 
\deleq{\[N_x(t) = \frac{1}{|X(t)|}\]}
The spectral consequences of this follow from the convolution theorem, according to which multiplying two signals in the time domain is equivalent to convolving them in the frequency domain, and vice versa.\del{, hence:}
\deleq{\begin{eqnarray}
\label{EQ_ConvTheorem}
	 \tilde{\Phi}_x(\omega)
	 = \int{
	 		\tilde{X}(\omega - \eta) 
			\tilde{N}_x(\eta)
			\mathop{d\eta}
			} 
\end{eqnarray}}
As a result, the spectrum of $\Phi$ is that of the original signal smeared with the spectrum of its inverse envelope, creating a form of spectral leakage bias. 

One worrisome aspect of this is that the inverse envelope tends to have a very broad spectrum because it contains large spikes wherever the envelope approaches zero; as a result, amplitude normalization tends to smear spectral energy across a wide swath of the spectrum. 
\del{The spectral properties of the signal envelope that give rise to this effect are considered in Appendix }
The effect of such spectral smearing can be appreciated most easily in phase ``jumps'' or ``slips'' within the normalized signal.
An example of this effect is shown in Fig. \ref{fig:beat_examp} for an amplitude modulated signal composed of the sum of a 10 Hz and a 12 Hz sinusoid, giving an 11 Hz signal modulated at 2 Hz. 
The inverse envelope contains energy at all harmonics of the modulation frequency, as a result of which the normalized signal also contains energy at harmonics of 2 Hz in addition to 10 Hz and 12 Hz. 
These new harmonics, which were not present in the original signal, may create spurious phase locking if they overlap with components of a compared signal. 

\del{More generally, the squared envelope of an analytic signal is obtained from the product of the signal with its complex conjugate; for an analytic signal occupying a bandwidth of  $\Delta W$, the corresponding convolution creates a spectrum contained in the interval $\left[-\Delta W, \Delta W\right]$. 
Whereas all even powers of the envelope are similarly band limited, odd, negative and non-integer powers are not (see Appendix 
for a more detailed discussion and proof). Any transformation applied to the envelope therefore produces an infinitely extended spectrum except in circumscribed cases involving only a finite number of even powers. }


\subsection{Some Nasty Statistics, or, the Futility of Filtering}
\label{NastyStats}

If the example in Fig. \ref{fig:beat_examp} seems contrived to produce broad spectral artifacts, one may consider the somewhat more relevant cases of random noise. 
For an analytic signal whose real and imaginary parts are independent and Gaussian, amplitude follows a Rayleigh distribution \cite{krishnamoorthy2015handbook}. 
For this distribution one finds that the inverse amplitude has a finite expected value,
\[\left<|X|^{-1}\right> = \frac{1}{\sigma^2}\sqrt{\frac{\pi}{2}}\]
but unbounded variance, 
\[\left<|X|^{-2}\right> \rightarrow \infty\]
meaning in practice that sample variance for the inverse amplitude will increase with sample size as the number of intermittently appearing large outliers grows, creating a signal whose energy comes increasingly to be dominated by such outliers, whose spectrum therefore becomes broader as its duration increases. 

It is natural to suppose that bandpass filtering does away with this problem by smoothing out the dips in amplitude responsible for spectral broadening of the inverse. 
Casual inspection of the signal  might lead one to take this point for granted, as a noisy signal initially full of phase jumps and amplitude kinks looks smooth and oscillatory after filtering.
Unfortunately, any such improvement is an illusion: filtering only decreases the effective sample size of the signal, thereby decreasing sample variance of the inverse amplitude, but it does not change the underlying distribution, which remains as poorly behaved as before.
To see why this must be true, one need only note that decimating a stationary narrowband Gaussian signal to a sampling rate that matches its bandwidth should yield nearly independent Gaussian samples\del{, without changing}\add{; but decimation does not change} the ergodic distribution of sampled amplitudes.  

\add{
A similar conclusion may be reached in an entirely different way\cite{picinbono1997on}: consider that the autocorrelation of the spectrum of a signal is the Fourier transform of the square of the signal's time envelope. 
The constant envelope of $\Phi$ entails a spectrum composed of a single delta spike at 0 Hz. 
If a discrete finitely sampled spectrum is band-limited, its autocorrelation must contain a nonzero value at the lag separating the outermost nonzero samples.
This precludes the possibility that a band-limited spectrum of a finite signal can generate the required autocorrelation, unless the bandwidth is infinitesimal, in which case the signal must be a pure sinusoid. 
This argument leaves no escape from the conclusion that amplitude normalization causes spectral broadening of the normalized signal, except in trivial cases.
}  


\del{It's worth dwelling on this point for a moment, because conventional wisdom holds that phase jumps, which contribute to spectral broadening of the normalized signal, are a symptom of an insufficiently narrow signal bandwidth. 
But this is not so; any apparent improvement that results from bandpass filtering is purely an artifact of the relationship between sample size and  variance for a random variable with bounded mean but unbounded variance 
Of course, this is not to say there are no good reasons to filter (some of which we will consider later), rather one should not be fooled into thinking that filtering leads to any inherent improvement in spectral properties of the estimator. 

%

In summary, amplitude normalization is destined to cause spectral broadening of the normalized signal under a wide range of conditions. 
The problem is compounded by the unsightly distributional properties of the inverse amplitude, which, among other things, creates some arbitrary dependence on signal duration and filtering bandwidth by way of effective sample size.

These points have important implications for measures of spectral dependence derived from amplitude normalized signals. 
Moreover, because the distribution of resulting phase vectors stands in one-to-one correspondence with the distribution of phase angles, the spectral leakage bias resulting from this treatment of the signal will also extend to measures derived directly from the empirical distribution of phase angles, even those that avoid any explicit normalization. }

\section{Decentering Bias}
\label{section:decbias}

\begin{figure} 
\includegraphics{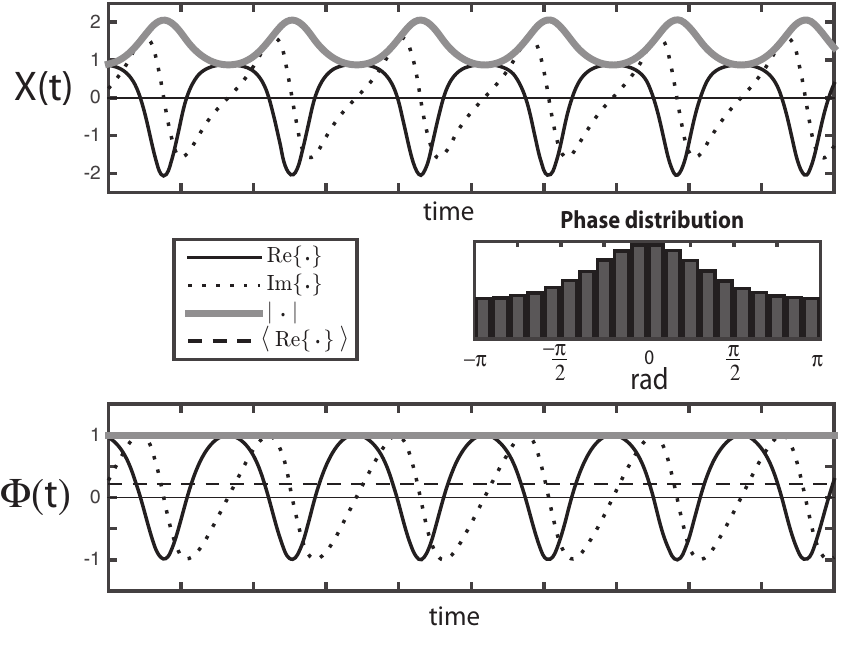}
\caption{\label{fig:phase_dist_examp} 
Example of decentering bias.
Before amplitude normalization a test signal with non-uniform phase distribution (\emph{inset}) has  a mean of zero (\emph{top panel}).
 After normalization (\emph{bottom panel}) the mean is offset from zero (\emph{dashed line}). 
This effect can be understood as an instance of spectral leakage from the original signal bandwidth to 0 Hz following normalization. 
\add{Before normalization,} the analytic amplitude (\emph{gray line}) is modulated with the same period as the original signal.  
This fact induces the 0 Hz component following multiplication by the inverse amplitude. 
}
\end{figure}

Spectral leakage from amplitude normalization increases the opportunity for bias in measures of phase locking by spreading energy in the Fourier domain, but this effect is not sufficient \del{in}\add{by} itself to create such bias. 
We will now consider a specific example of how spectral leakage may cause the appearance of a phase relationship between signals that are, in fact, entirely independent.
As touched on in the discussion of Eq. (\ref{EQ_cspec_est}), cross-spectral estimation generally \del{doesn't} \add{does not} involve mean centering because the input signals are band-limited by assumption and so should already have an expectation of 0. 
Clearly, spectral leakage means this condition may no longer apply following amplitude normalization.  

An instance of this effect was recently highlighted by van Driel et al. \cite{van2015phase}, who describe it as resulting from non-uniformity of the distribution of phase angles in the input signals. 
If this distribution is not at least radially symmetric, any non-uniformity will cause a deviation of the expected phase vector from the origin.
\del{The problem is therefore essentially the same error that results from correlating two real variables without subtracting their means: For independent $X$ and $Y$, $\mathrm{E}[XY^*] = \mathrm{E}[X]\mathrm{E}[Y^*]$, and so} 
One obtains a nonzero value in the absence of any true correlation if both expectations are nonzero. 
One solution may seem obvious at this point: subtract the mean phase vector from at least one of the signals before computing any statistics related to phase locking.
From our Fourier domain perspective, this correction amounts to applying a \del{minimally}\add{maximally} high-pass filter, eliminating only the 0 Hz component. There are immediate problems with \add{this} solution, however, which will be considered in the following section.

First, to understand how energy comes to be leaked to 0 Hz, consider that a peak in the distribution of a signal's phase implies the presence of amplitude and frequency modulation whose spectrum overlaps with that of the original  signal (Fig. \ref{fig:phase_dist_examp}). 
This is easiest to understand in terms of frequency modulation, as a peak at a given phase value means that the progression of phase in the signal must linger around the peak value, implying a drop in average instantaneous frequency.
For the signal to be zero-mean, such a dip must be accompanied by a drop in amplitude to compensate for the corresponding bias in the direction of phase. 
These spectral features in the amplitude carry over to its inverse, and normalization therefore induces a 0 Hz spectral component through the cancellation of the dominant component in the original signal and the matching (phase-locked) component in amplitude modulation. 

Decentering bias therefore occurs whenever the \del{direction}\add{phase} and magnitude of the underlying signal are correlated.
\add{This can also be shown from the fact that the covariance of amplitude and phase vector yields the expectation of the original signal, which here is assumed to be 0:
\begin{align}
\mathrm{E}\left[|X|\Phi_X\right] =  \mathrm{E}[X] = 0
\end{align}
so that the two are second-order independent only if phase is centered:
\begin{align}
\mathrm{E}\left[|X|\Phi_X\right] = \mathrm{E}\left[|X|\right]\mathrm{E}\left[\Phi_X\right] \quad \mathrm{iff} \quad   \mathrm{E}\left[\Phi_X\right] = 0
\end{align}
} 
For this to happen in a stationary signal, the signal bandwidth, which governs the spectral content of amplitude fluctuations, generally must approach the spectral range of the signal, that is, $\Delta W \approx f_1$, where $f_1$ is the lower bound of the signal spectrum; however, because spectral leakage \add{from amplitude normalization} may be infinitely extended, this condition should be taken as a rule of thumb rather than a guarantee. \add{The example of two interfering sinusoids  in } Fig. \ref{fig:beat_examp} gives a \add{simple} case in which spectral leakage extends to 0 Hz, even though bandwidth condition is met.  

\del{When phase locking is computed only at selected time points In this setting, another another cause of amplitude-phase correlation is signal non-stationarity. 
In the context of electrophysiology, for example, such a dependence may be caused by an evoked potential.
A basic example of how amplitude-phase correlation produces decentering bias in both cases is presented in Fig. 
, which also compares its effect on PLV and some alternative measures considered in following sections.}

It should be noted here that in spite of the examples considered in the previous sections, \add{stationary} Gaussian signals are not susceptible to this type of bias because they are completely characterized by their first two moments (mean and cross-covariance). 
Any dependence between the signal envelope and phase must involve moments of order 3 or higher. 
For this reason, if the originating signals are  strictly Gaussian, pure phase-locking measures must still be directly related to second order statistics, which has recently been confirmed by Aydore et al. \cite{aydore2013}. 
But many if not most biological signals are manifestly non-Gaussian; this fact is exploited by widely-used blind-source separation techniques, which equate sources with maximally non-Gaussian signal components \cite{vigario2000}. 
Here it will likewise be taken for granted that the signals in question are non-Gaussian.

\subsection{The Problem with Recentering}
\label{section:RecenteringProblem}
Centering the normalized signal removes the bias resulting from an off-center mean. 
It has been suggested that this correction effectively serves to uniformize the phase distribution, implying that it deals more generally with problems related to non-uniform phase \cite{van2015phase}.
Such a conclusion is premature\del{, however,} because it neglects another important consequence of recentering: following mean subtraction, phase vectors no longer have uniformly unit length. 
It is therefore mistaken to suppose that mean centering of the normalized signal implies a balancing of the distribution of phase angles.
Instead, the correction depends in part on a re-injection of some of modulation that was removed through amplitude normalization. 

The centered signal is:
\begin{equation}
\label{EQ_centeredphi}
\begin{split}
	\Phi^*(t)&= h(t)e^{i\phi^*(t)} \\
		    &= e^{i\phi(t)} - |\mu_\phi|e^{i\theta_\mu}
\end{split}
\end{equation}
where the amplitude-normalized signal before centering is 
$\Phi(t) = e^{i\phi(t)}$ 
and the centering term is given as 
$\left<\Phi(t)\right>=|\mu_\phi|e^{i\theta_\mu}$. 
It is easy to see that this results in an amplitude modulated signal with envelope:
\begin{equation}
\label{EQ_centered}
	 h(t) = \sqrt{1 + |\mu_\phi|^2 - 2|\mu_\phi| \cos\left(\phi(t) - \theta_\mu\right) }
\end{equation}
This mean-centered signal  is no longer the pure representation of phase we had hoped to obtain, and we find ourselves back in the position from which we started.

\subsection{Corrections to Recentering}
\label{section:MC_correction}

%

\begin{figure*} 
\includegraphics{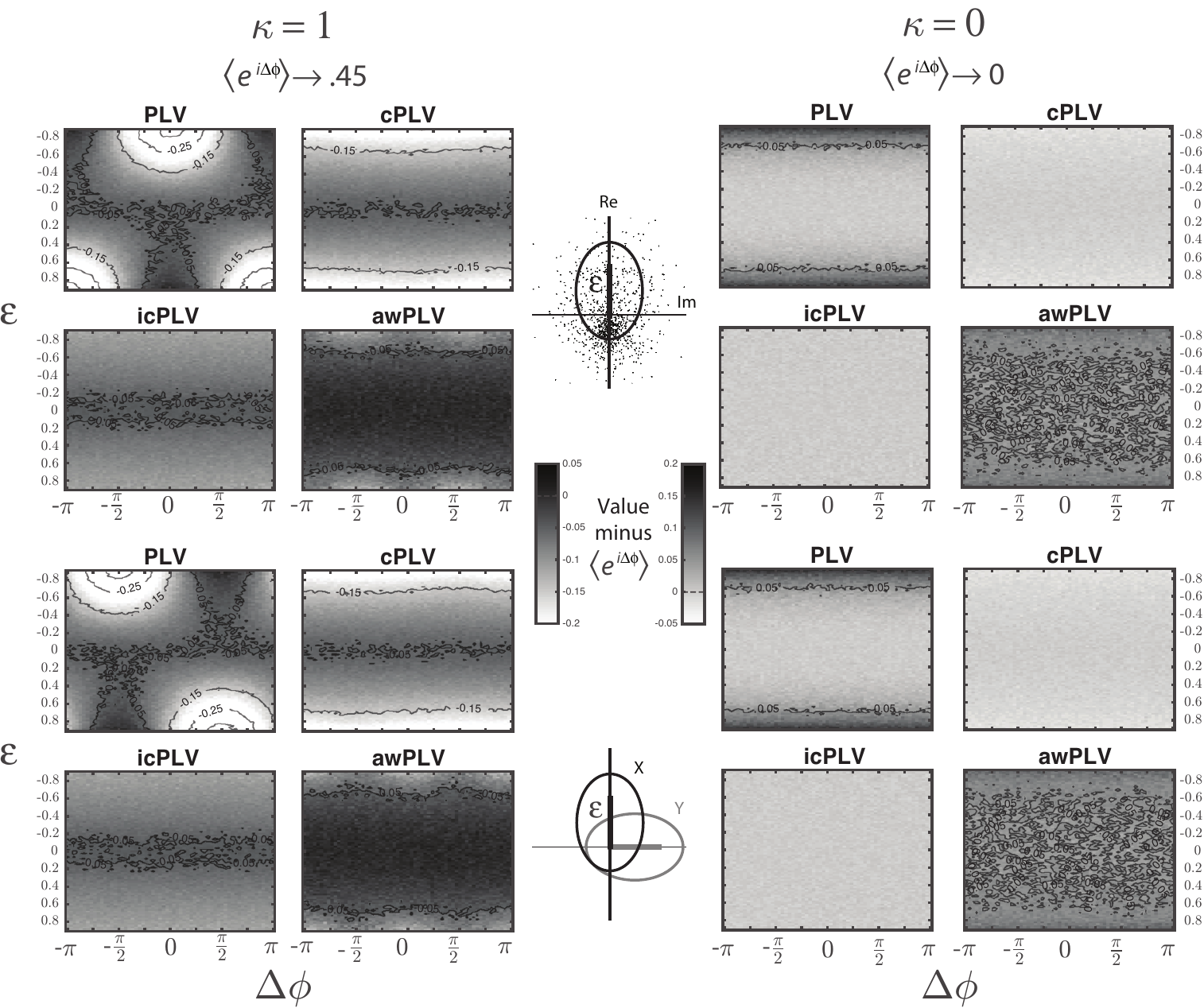}
\caption{\label{fig:PL_comparison} 
Comparison of decentering  bias in different measures. 
Test data are generated by skewing the magnitude  of independent Gaussian samples drawn from two processes exhibiting phase dependence. For both processes, amplitude is skewed according to an ellipse (example given in upper inset panel), with the degree of skew related to eccentricity ($\epsilon$). 
The phase differences $(\Delta\phi)$ between samples from each process follow a von Mises distribution with scale parameter, $\kappa$.
Compared measures are phase-locking value (PLV), centered phase-locking value (cPLV), iteratively centered phase locking (icPLV) and amplitude-weighted phase locking (awPLV).
 \emph{Left columns}, $\kappa=1.0$, giving expected unbiased PLV of 0.45; \emph{Right columns}, $\kappa=0.0$, giving expected unbiased PLV of 0 (i.e. no phase locking). 
The directional dependence of amplitude results in decentering bias in measures of phase locking that normalize by magnitude; this bias is shown as the deviation from the proper phase locking value for the von Mises distribution used to generate phase differences. 
Displayed values are  averaged over 100 draws of 100 samples each and shown as a function of $\epsilon$ and $\Delta\phi$. Four cases are addressed: \emph{Top rows}: amplitude skew aligned in the same ($\epsilon>0$) or opposite ($\epsilon<0$) directions; \emph{Bottom rows}: skew at right angles.
 In the presence of actual phase locking (\emph{left columns}), PLV and cPLV underestimate phase locking when the difference matches the skew directions. 
 In the absence of phase locking (\emph{right columns}), PLV overestimates phase locking at large $\epsilon$ (note different color scales for left and right columns). Other measures are not affected. The differences in awPLV are a consequence of increasing small sample bias rather than decentering (see section \ref{section_EDOF} in the text).
 }
\end{figure*}

The foregoing example illustrates the whack-a-mole character such corrections often assume in the spectral domain, where compensating for one artifact introduces others elsewhere. 
\del{The compliance of cross-spectral measures in this respect is one of their most appealing qualities. We will consider how to carry some of these qualities over to measures that avoid some of the ambiguities of phase locking.
But first, we take a brief detour to point out a correction for the recentering procedure proposed van Driel et al, which brings their debiasing procedure closer to its original intent.} 
\add{To preserve the desired properties of the phase vector, we briefly point out two improvements to mean-subtraction. 

\subsubsection{Iterated recentering} 
\label{section:iterated_recentering}

The first of these} \del{This correction} simply repeats recentering and normalization through multiple iterations:
\begin{equation}
\label{EQ_iter}
	\Phi_{\mathrm{ic}}^{(n+1)}(t)= \frac{
					 	\Phi_{\mathrm{ic}}^{(n)}(t) - \mu^{(n)}_\phi  
					 }{
					 	h^{(n)}(t)
					 }						 	
\end{equation}
We have observed this to converge rapidly on a phase representation with the desired properties: unit amplitude and zero mean. 
The resulting distribution of phase is radially symmetric but not generally uniform.
\del{The outcome of this procedure is a representation of signal phase with zero mean, which also correctly preserves unit amplitude of the phase vectors.  
An example is shown in 
using the test signal from
The success of the algorithm is reflected in the evening  out of the distribution of phase angles, which ends up not perfectly uniform, but symmetric in a way that gives a zero mean phase vector.  
also shows that the modulus of the signal mean after renormalization converges exponentially to zero.  
However, iterated or not, there is no reason to expect that mean centering will improve any of the other spectral artifacts noted above. 
The next section describes conditions which will cause recentering to fail.}

\add{
\subsubsection{Phase Uniformization}
\label{section:phase_uniform}
The second correction enforces a strictly uniform distribution by applying a suitable transformation to the phase angle  \cite{kralemann2008phase}. 
This is easily done by passing the angular representation of phase through its cumulative distribution over $[0,2\pi]$
\[\phi_\mathrm{u}(t) =  2\pi\hat{F}\left(\phi(t)\right) \]
Such a correction is a more general fix for any problem that might arise from a lack of uniformity. 
A possible drawback in some settings is the need to estimate the cumulative distribution function, $\hat{F}$, which is a more complex problem than recentering, involving more degrees of freedom. 
It may therefore not suit cases in which a reliable estimate is difficult to obtain, such as with small samples.

Most importantly, neither of these approaches addresses the general problem of spectral bias. 
They only correct the problem as it affects mean phase.   
We next consider how and when such corrections fail.}
 
%

\section{Bias and Correlated Amplitude Modulation}
\label{section:ampmod}
We have just seen a concrete example of how spectral leakage might affect measures of phase locking through decentering of the amplitude normalized signal. 
We also reviewed some simple techniques to correct that particular incarnation of bias. 
Next we will describe conditions under which decentering bias can escape any correction designed to recenter the distribution of phase. 
The key to understanding how this comes about lies, once again, in the spectral effect of amplitude modulation.

\begin{figure}
\includegraphics{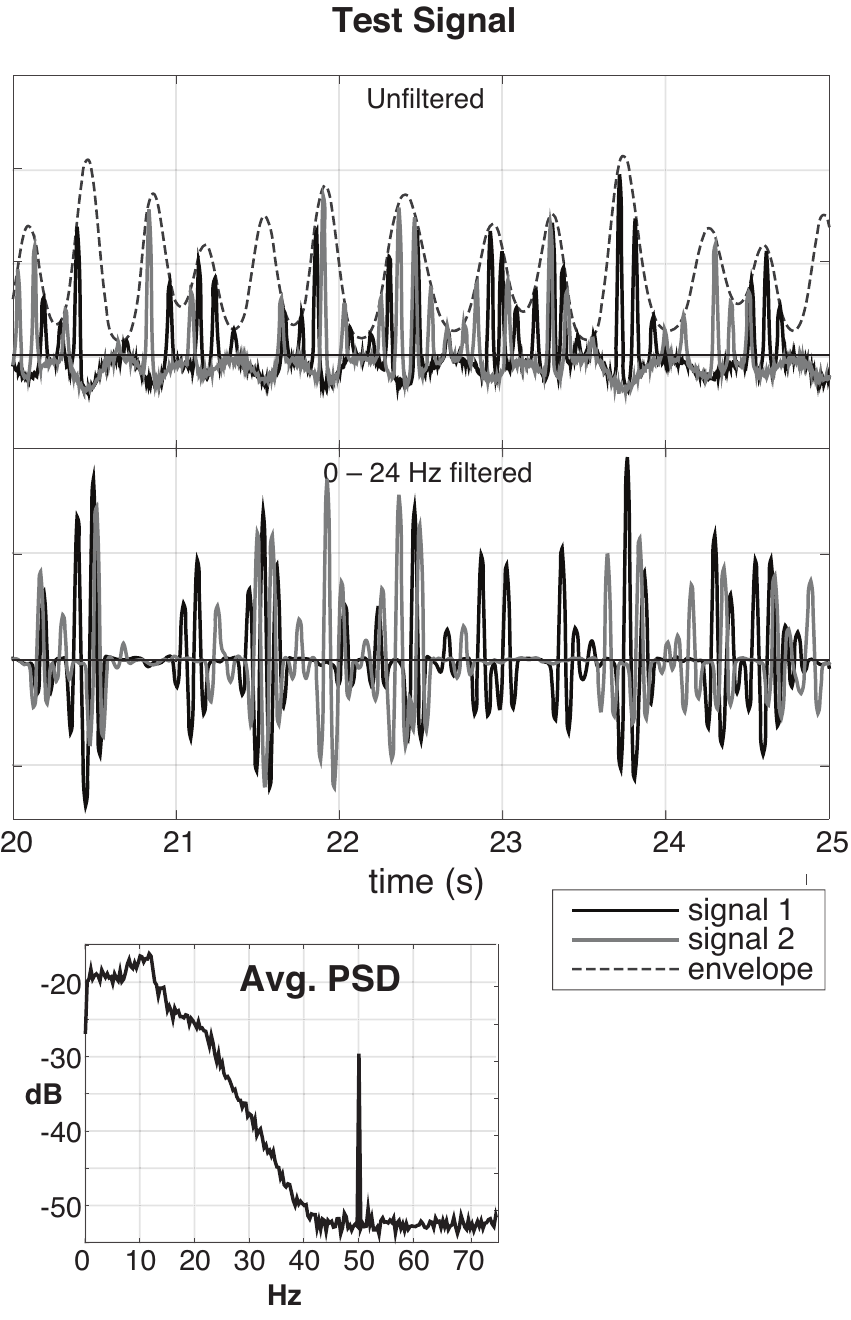}
\caption{\label{fig:test_signal} 
\add{Test signal used to illustrate shifted decentering bias in Fig. \ref{fig:rand_gauss}.}  \emph{Top panel:}  Five second example from a random test signal.
Test signals contained random and independent uncorrelated Gaussian transients modulated by a common 0--6 Hz signal.
The modulating signal (dashed line) was obtained by squaring 1.5--3 Hz filtered white noise with mean of 1 and standard deviation of 0.5 after filtering, to which a further positive 0.25 offset was added after squaring. Power spectral density is shown in the bottom panel.
The common envelope was applied to 100 random uncorrelated trains of 25 ms std. width Guassians with center times determined by positive peaks in 8--12 Hz filtered white noise. 
A common 50 Hz pure sinusoid served as a positive comparison. 
Total signal duration was 80 s.
\emph{Middle panel:} Signal after application of an analysis filter centered at 12 Hz. 
The analysis filter applied a frequency-domain cosine window over 0--24 Hz. 
The presence of the common low-frequency amplitude modulation creates spurious phase locking in the range of the uncorrelated train of gaussians at 12 Hz (see Fig. \ref{fig:rand_gauss}).
}
\end{figure}

Modulating (i.e. multiplying) a signal, $X(t)$ by a pure complex sinusoid, $e^{i\omega_c t}$, has the effect of shifting the entirety of the signal's spectrum over by an amount equal to the frequency of the sinusoid. 
This follows directly from the convolution theorem. 
\del{and from the fact that the Fourier domain representation of the sinusoid $e^{i\omega_c t}$ is a delta function ``spike'' at $\omega_c$, $\delta(\omega -\omega_c)$.}
\deleq{ 
\begin{equation}
\begin{split}
\label{EQ_ShiftTheorem}
	\mathscr{F}\left[X(t)e^{i\omega_c t}\right]
	 &= \int{
	 		\tilde{X}(\omega - \eta) 
			\delta(\eta-\omega_c)
			\mathop{d\eta}
			}\\
	&= \tilde{X}(\omega - \omega_c)
\end{split}
\end{equation}}
The result, $\tilde{X}(\omega - \omega_c)$, is therefore in every respect identical to the original spectrum of $\tilde{X}(\omega)$, except for having been shifted along the frequency axis by $\omega_c$.

This property brings up an important point that will prove conceptually useful: the invariance of the spectrum under frequency translation carries over to all measures of phase locking we have encountered so far. 
This follows from the fact that the modulating term $\omega_c t$ enters into the complex argument (phase angle) of the modulated signal\del{.
It} and will therefore cancel when computing the phase difference between two signals that have been modulated by the same sinusoid, resulting in the same outcome as if no modulation had been applied to either. 
\del{Likewise, any such modulation cancels in computing the envelope for each signal. 
In short, modulation with a complex sinusoid has no effect on any measure of phase locking derived from the phase difference between signals that have been modulated in the same way.}
It should also be apparent that this property extends more generally to any complex-valued modulating signal with unit amplitude, $e^{i\beta(t)}$. 
\del{How this relates to modulation by real signals will become clear shortly.
The property of invariance under translation has important implications for the decentering bias of modulated signals.}  
Modulation of the original signal with a complex unit-amplitude \del{signal}\add{function} will \add{therefore} cause some of the 0 Hz component of the normalized signal, responsible for decentering bias, to shift to modulating frequency(-ies). 
In this scenario, it is possible for the mean phase vector to be zero and the distribution of phase angles perfectly uniform in the input signals. 
Yet, if the two signals under comparison share amplitude modulation, the cancellation of the common component in the argument reestablishes decentering bias.

\begin{figure*} 
\includegraphics{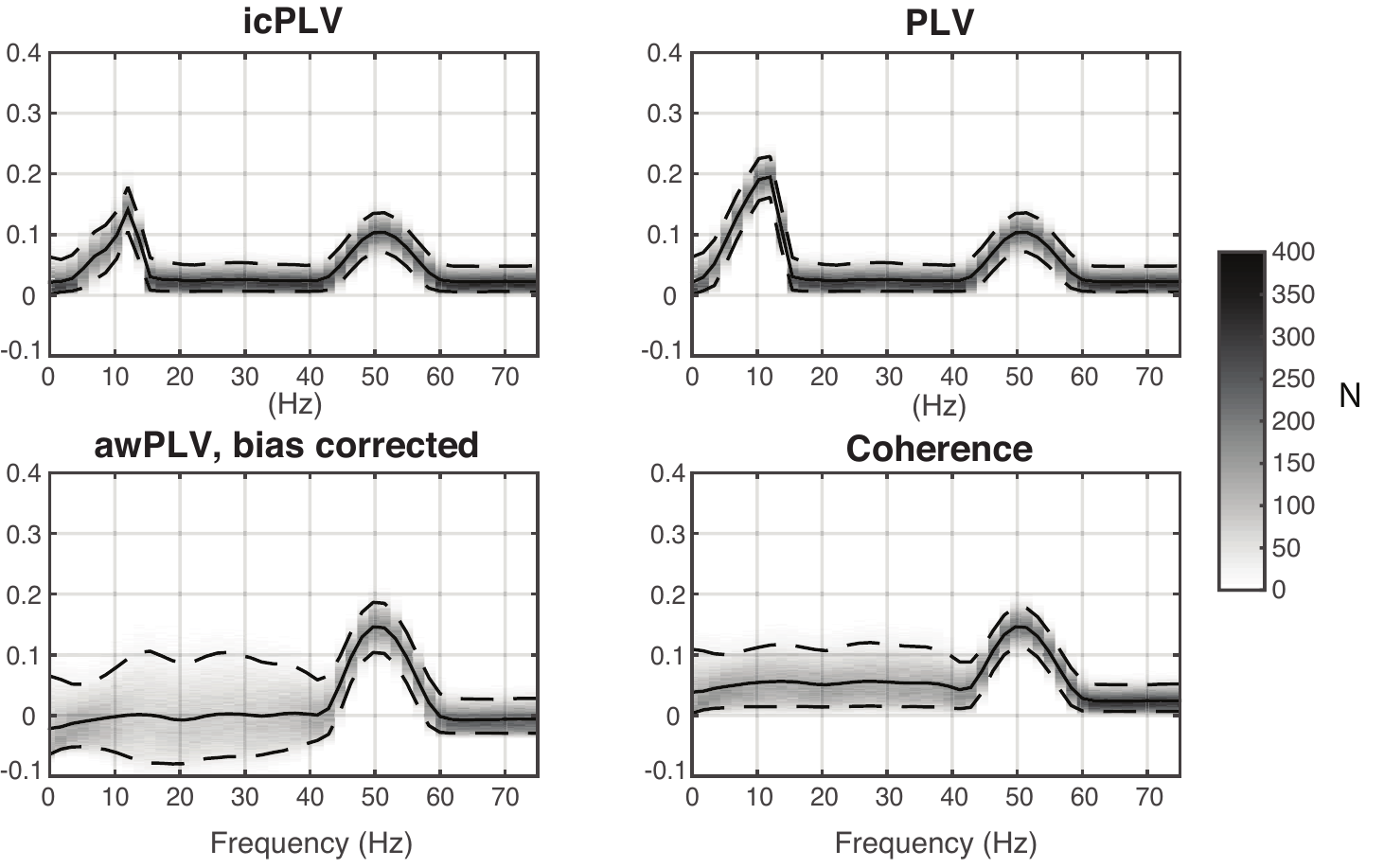}
\caption{\label{fig:rand_gauss} 
Spectral bias in four measures of phase dependence applied to test signals exemplified in Fig. \ref{fig:test_signal}.
Phase-locking measures were computed between all 10,000 pairs of signals drawn from two sets of 100, and the figure color scale indicates the distribution of values for each frequency bin. 
Lines show the median, 5th and 95th percentiles of these distributions.  In all cases, the peak at 50 Hz reflects the 50 Hz positive comparison signal component.
\emph{Top panels:} Mean-centered (icPLV) and conventional phase-locking values (PLV) both show spurious phase locking around 12 Hz, in the frequency range of the uncorrelated train of gaussians, as a result of a common 1.5--3 Hz modulating envelope. 
Mean centering creates relatively greater suppression of the bias at low frequencies (\emph{top left panel}), which reflects the contribution of the unmodulated constant component of the envelope. 
\emph{Bottom panels:} awPLV and coherence are not affected by this bias. 
Measures were computed within overlapping cosine-windowed 16 Hz frequency bands using the DBT method (Kovach and Gander, 2016).  
}
\end{figure*}

\del{This basic insight may be carried over to}\add{The relevance of this to} real-valued \del{case}\add{signals becomes clear} by constructing the modulating signal from a zero-mean analytic signal $B(t)=\add{|B(t)|e^{i\beta(t)}}$ representing amplitude fluctuations around a constant offset $\mu_b$:
\begin{equation}
\label{EQ_AMsignal}
\begin{split}
	b(t) &=|B(t)|\cos\left(\beta(t)\right) + \mu_b\\
	&= \frac{1}{2}|B(t)|\left(e^{i\beta(t)}+e^{-i\beta(t)}\right)	+ \mu_b       
\end{split}
\end{equation}
\add{Provided the spectrum of $b(t)$ is confined to a lower range of frequencies than the signal it modulates, $x(t)$, the Hilbert transform preserves the amplitude modulation with $b$ (Bedrosian's theorem)  \cite{bedrosian1963product}:
\begin{align} 
	\label{EQ_bedrosianThm}
	\mathscr{H}\{b(t)x(t)\}=b(t)\mathscr{H}\{x(t)\}
\end{align}
Consequently, the analytic representation of the modulated signal is 
\begin{align}
\label{bx2bX}
\begin{split}
&A_{bx}(t)e^{\phi_{bx}(t)}= b(t)X(t)\\
=&\left[\left|B(t)\right|e^{i\beta(t)}+\left|B(t)\right|e^{-i\beta(t)}+\mu_b\right]X(t)
\end{split}
\end{align}  }
From Eq. (\ref{bx2bX}), one can see that modulation with a low-frequency real-valued sinusoid (e.g. $b(t)=\cos(\omega_c t) $, $\mu_b=0$) has the effect of reduplicating and summing the original spectrum, translating one copy by $+\omega_c$ and the other by $-\omega_c$ along the frequency axis. 
More generally, a smeared out version of the original spectrum will be translated in both directions by an amount reflecting the spectral range of $B$, creating ``sidebands''.
 If $\mu_b$ is nonzero, then a third copy at the original location, the ``baseband'', will reflect the presence of an offset (non-vanishing 0 Hz component) in the spectrum of $b$. 
 \del{This case is illustrated in Fig.}

\del{What} \add{The} consequence this will have for \del{any}measures of phase locking\add{ depends on the analysis filter used to obtain the relevant statistics.} \del{depends on which parts of the smeared out spectrum are covered by the frequency window  used to obtain the relevant statistics, here referred to as the ``analysis filter.'' }
\add{If the passband covers the full spectral range of the modulated signal, $b(t)X(t)$, and if $b$ is positive, the distribution of phase remains unaffected because the modulation is retained within the amplitude of the result. 
If the passband instead isolates one of the side bands, $|B|e^{ \pm i\beta}$, then the effect of amplitude modulation enters into the phase of the result by way of the complex argument. }

Appendix \ref{appendix:AMspect} reviews three cases: (1) an analysis filter whose passband covers the entire spectrum of the modulated signal (broadband case); (2) a passband that covers only one of the shifted sidebands (sideband case); and (3) a passband that covers the unshifted band induced by any nonzero offset of the modulating signal, as occurs when the modulating signal is strictly non-negative (baseband case).  
\del{It is shown there that}\add{In all cases, pure phase-locking measures remain \del{entirely} unaffected by the modulating signal because any \add{influence on} the phase of the separate signals cancels from the difference}. 
\del{This comes as no surprise, given that eliminating the influence of amplitude motivates  such measures in the first place.} 
\del{But it also}\add{This} means\add{, however} that any decentering bias in the unmodulated case remains in effect after modulation, even though the distribution of phase may be uniform within each side band. 
The off-center mean may be obscured as a result of phase dispersion introduced by amplitude modulation, making recentering ineffective.
\del{If the analysis filter isolates individual bands, recentering still suppresses bias in the baseband, but not the side bands.}
This effect is illustrated in Fig. \ref{fig:rand_gauss}, which also compares some of the alternative measures discussed in the following section,  using test signals described in Fig. \ref{fig:test_signal}. 

\add{More generally, it is not necessary to assume that the bands induced by amplitude modulation are fully separated by the analysis filter. When the bands overlap so that the modulated signal lacks distinct side bands or the analysis filter otherwise provides an incomplete separation, the filtered signal contains an unequally weighted sum of components arising from each band.
In such cases, the effect of amplitude modulation on the phase of the result is will be intermediate to the side- and broadband cases, depending on relative weighting, so that recentering is still incompletely effective. }

\section{Remedies}
\label{Remedies}
We have seen that measures of dependence taken directly from the cross spectrum are not affected by the potentially severe forms of spectral leakage produced by amplitude normalization (Fig. \ref{fig:rand_gauss}). 
But the canonical measure in this category, coherency, has other drawbacks, which diminish its attractiveness as a proxy for phase locking; in particular, it combines phase locking and amplitude correlation in a way that complicates interpretation. 
In the following we consider how the simple average used by PLV can be replaced with a weighted average. 
A variant of this approach, amplitude-weighted (awPLV), avoids  amplitude normalization altogether, instead retaining the original cross-spectral amplitude weighting but with a rescaling that overcomes some of the interpretational difficulties of coherence. 
These measures avoid any effects of spectral broadening that follow from amplitude normalization. We also briefly consider more general extensions of this scheme to other types of weighted averages. 

\subsection{Amplitude-weighted Phase Locking}
\label{section_awPLV}
\add{The covariance of two analytic signals that are perfectly phase locked, meaning $\Delta\phi_{xy}=\phi_x(t)-\phi_y(t)$ remains constant, is given by:
\begin{equation}
\label{EQ_csindep}
\begin{split}
E\left[S_{xy}\right] =& \int{E\left[A_x(t)A_y(t)e^{i\Delta\phi_{xy}(t)}\right] \mathop{dt}}\\
=& e^{i\Delta\phi_{xy}}\int{E\left[A_x(t)A_y(t)\right] \mathop{dt}}
\end{split}
\end{equation}
One might therefore consider a measure of ``pure'' phase locking that factors out the influence of amplitude covariance simply by normalizing the cross spectrum with $\int{E\left[A_x(t)A_y(t)\right] \mathop{dt}}$.
Equivalently, such a measure is obtained from the normalization of coherency by the uncentered amplitude correlation in the presence of phase locking.}
\del{The alternative measure proposed here, amplitude-weighted phase locking (awPLV), gets around this problem very simply, by normalizing coherence with amplitude correlation.}
\begin{equation}
\label{EQ_awPLV}
\begin{split}
     \Upsilon_{xy} &= \frac{
     				  \left<XY^*\right>
				  }{
				  \sqrt{\left<|X|^2\right>\left<|Y|^2\right>}
				  } 
			\left/			  
					  \frac{
					  		\left<|X||Y|\right>
						}{
							\sqrt{\left<|X|^2\right>\left<|Y|^2\right>} 
							}
			\right.\\
	     &=  \frac{
     				  \left<XY^*\right>
				  }{
				 \left<|X||Y|\right>
				 }
	   =  \frac{
     				  \left<A_xA_ye^{\Delta\phi_{xy}}\right>
				  }{
				 \left<A_xA_y\right>
				 }
 \end{split}
\end{equation}
The result of this normalization has a natural interpretation as a weighted-average phase coherence:
\begin{equation}
\label{EQ_awPLV2}
\begin{split}
    \hat{ \Upsilon}_{xy} &= 
     				  \sum_t{w_te^{\Delta\phi_{xy}[t_i]}}				 
 \end{split}
\end{equation}
with weights given by normalized amplitude products: 
\begin{equation}
\label{EQ_awPLVwgt}
 	   w_t = \frac{
	   		A_xA_y
	   		}{ 
			\sum_t{A_xA_y}
			}				 
\end{equation}
The meaning of this measure is therefore much more closely aligned with that of phase-only measures, despite its close relationship to coherence. 
Like PLV, its magnitude becomes 1 if the signals are perfectly phase locked, regardless of how the amplitudes are related.

This measure will differ from standard PLV in three ways; first, as already noted, it avoids spectral biases of the type reviewed above. 
Second, like coherence, awPLV weights the estimate towards samples with relatively greater amplitude, and therefore will diverge from PLV if the phase difference exhibits dependence on signal amplitudes. 
Third, as reviewed in the following section, statistical properties of awPLV will depend on effective sample size, which is always smaller for a non-uniformly weighted average than for an unweighted average.

\subsubsection{Effective Sample Size}
\label{section_EDOF}
For a non-uniformly weighted average, sample size is no longer an accurate reflection of the degrees of freedom in the average\del{. Statistical degrees of freedom are}\add{, which is} more accurately reflected by\del{ a measure that accounts for the variable contribution of samples,} the effective sample size, $\nu$. For independent samples, effective sample size is approximated as 
\begin{equation}
\label{EQ_awEDF}
    	\nu= \frac{
     				  \left(\sum_t{w_t}\right)^2
				  }{
				  \sum_t{w_t^2}
				  } 		
\end{equation}
In general, $\nu \le N$. In the absence of phase locking, the expected value of PLV is related to sample size; this small-sample bias is given by the inverse square root of sample size, $1/\sqrt{N}$ (assuming independent samples). Small-sample bias is approximated for awPLV in the same way, substituting effective sample size:
\begin{equation}
\label{EQ_awEDF}
    	\beta = \frac{ 1 }{
				  \sqrt{\nu}
				  } 		
\end{equation}
Using this relationship one may define a bias-corrected measure, which is centered \del{on }\add{at} zero in the absence of phase locking and otherwise uses the full dynamic range between 0 and 1:
\begin{equation}
\label{EQ_awPLVBC}
\begin{split}
    \hat{ \Upsilon}^*_{xy} &= 
     				  \frac{ \hat{ \Upsilon}_{xy} - \beta}{1-\beta}		 
 \end{split}
\end{equation}
This correction addresses another shortcoming of standard coherence estimates, which is that the degree of bias varies across frequency, depending on the amplitude distributions within the respective analysis bands \cite{lowet2016quantifying}, potentially creating the false impression of frequency-dependent changes in coherence.

While inflated small-sample bias is clearly an important consequence and possible disadvantage of non-uniform weighting, the apparent loss of statistical power in awPLV and coherence relative to PLV is very likely to be offset by the fact that samples with amplitude approaching zero, for which phase is either poorly defined or unlikely to reflect any signal beyond the noise floor, are granted equal weight in PLV.
To the extent that phase locking is associated with the amplitudes of the carrier signals, awPLV will therefore  tend to improve the signal to noise ratio.
On the other hand, for a signal contaminated with high-amplitude transients, such as movement artifacts, awPLV will perform poorly, raising the importance of careful artifact rejection.

For a bandpass-filtered signal, Eq. (\ref{EQ_awEDF}) is clearly not valid when summing over contiguous samples because filtering creates correlations among samples over time, violating the assumption of independence. Using (\ref{EQ_awEDF}) will give a greatly inflated estimate of the effective sample size in such cases. One way to correct the problem is by downsampling the signal so that samples are more nearly independent; this is conveniently addressed by spectral estimation techniques that combine bandpass filtering with downsampling, adjusting sampling to the bandwidth of the analysis filter. We have recently described such an approach, the demodulated band transform (DBT) \cite{kovach2015demodulated}, which uses a frequency-domain implementation of complex demodulation \cite{bingham1967modern}.

Effective sample size is strictly valid only when weights are independent of phase, which may often fail to hold in the presence of phase locking. It is, however, a valid assumption under the null hypothesis of no phase locking. Null hypothesis tests on awPLV address the question\del{,} how likely \del{is} one \add{is} to obtain the observed value for two signals with the respective amplitude envelopes in the absence of phase locking. 

\subsection{Permutation Tests}
Statistical validation of phase locking frequently resorts to a surrogate null distribution generated by shuffling data over samples. 
The rationale is that shuffling abolishes any true phase locking while leaving biases intact.
The considerations reviewed here show the limits of this assumption. 
In the case of pure measures of phase locking, we have seen that bias may depend on correlations of the signal amplitudes. 
Clearly, shuffling destroys any amplitude correlation along with phase locking, meaning that the surrogate distribution will fail to account for any related bias. 
Although awPLV avoids spectral biases, it is not immune to this problem because shuffling may systematically change effective sample size, likewise ruining the validity of the comparison.
By increasing the variance of the product of amplitudes, positive correlations depress effective sample size; shuffling may therefore produce a systematic decrease in small sample bias, raising type I error in any test that relies on the comparison to shuffled data. 

There seem to be few perfectly  satisfactory options for addressing this problem. 
One is to shuffle phase independently of amplitude, but this creates distortions similar to those arising from amplitude normalization.
A better alternative might be to shuffle amplitude and phase in tandem but reduce (or increase) the number of samples in the averages drawn from the surrogate distribution in proportion to 
\begin{equation}
N^{(k)} = N\frac{\nu}{\nu^{(k)}}
\end{equation}
where $\nu^{(k)}$ is the effective sample size of the $k^{\mathrm{th}}$ resampling including $N$ samples. Such an adjustment at least removes the tendency for systematic changes in effective sample size, although it does not exactly recreate the amplitude distribution of the original signals. Another more complicated approach might be to stratify the permutation by amplitude, so that surrogate data maintain a similar amplitude profile. 

\subsection{Generalized Weighting}
\label{section:genweight}
One might also consider alternative weighting schemes other than simple amplitude weighting as used by awPLV; for example, Ehm et al. suggest discarding samples whenever amplitude drops below some threshold  \cite{ehm2009detecting}. In general, reweighting will take the form:
\begin{equation}
\label{EQ_GenWeight}
\begin{split}
    \hat{\Psi}_{xy} &= 
     				 \frac{
				 		 \int{ \Phi_{x}\Phi^*_{y}\mathop{dW(t,|X(t)|,|Y(t)|)}}		
					}{
					 \int{ \mathop{dW}}		
					 } 
 \end{split}
\end{equation}
The hat symbol, used here to indicate an estimator, is deliberately retained in (\ref{EQ_GenWeight}) because this notion of weighting also encompasses sampling of the signal at specific points in time within some window of observation. So that it applies to this range of cases, the notation in Eq.  (\ref{EQ_GenWeight})  expresses the measure as a Riemann--Stieltjes integral, which is a formality valid for both discrete atomic and continuous weighting.
In general, cross-spectral measures are optimal with respect to spectral leakage, so any form of weighting other than a rescaling of the amplitude product will create some additional spectral broadening.  
Nevertheless, reweighting may improve the measure in other ways, such as in moderating small-sample bias and other non-spectral-leakage-related biases imparted by amplitude weighting. 

Such reweighting is also implicit when samples are drawn from selected time points, as arises generally in the context of digital signal processing, but also more specifically when measuring phase locking at certain time delays, $\tau$, from some event, such as trial onset. 
This case is represented by the following estimator:
\begin{equation}
\label{EQ_trialWeight}
\begin{split}
    \hat{ \Psi}_{xy}(\tau) &= 
     				  \int{w(t)\Phi_{xy}\sum_i{\delta\left(t-t_i-\tau\right)}\mathop{dt}}	\\
				  &=\sum_i{w\left(t_i+\tau\right)\Phi_{xy}\left(t_i+\tau\right)}	 
 \end{split}
\end{equation}
for which weighting is the composite of a continuous weighting function $w$ and the subsampling represented by the delta train. 
To understand the spectral consequences of such subsampling, one may note that the Fourier transform of a train of delta functions with period $\Delta T$ is also a periodic train of spikes in the frequency domain, with period $2\pi/\Delta T$. One can observe that this form of ``windowing'' has the effect of reduplicating the original signal spectrum at frequency intervals of $2\pi/\Delta T$. 
If the period of the original train is less than the inverse bandwidth of the signal, expected cross-spectral estimates remain unchanged because the reduplicated spectra do not overlap. 
More generally, the cross-spectral estimate will be relatively unaffected if the minimum interval between samples is less than the inverse bandwidth of the signal.
This point simply restates the well-known Whittaker--Nyquist sampling theorem.

In event-related analyses the goal is to reveal nonstationarities associated with some event.  In this context, sampled points are widely spaced and the reduplicated spectral windows overlap, which implies convolutional smearing within the signal bandwidth. 
Any systematic change that results from this is not necessarily cause for concern, however, as it captures the nonstationarities that motivated the sampling in the first case. 

\section{Discussion}
\subsection{Some Advantages of Linearity}
\label{sec_linearity}
We have treated spectral broadening related to the nonlinearity of phase locking as a form of bias, though we also noted at the outset that this might fairly be considered an abuse of terminology. 
The nonlinearity of phase extraction is in fact faithful to the nature of analytic phase; what we have taken to calling bias is a genuine property of the signal, not the result of bias in the usual meaning of the word.
More accurately stated, the problem is a choice between approximations: whether to approximate phase locking with measures that preserve linearity or to treat nonlinear measures as approximately linear in our interpretation of them (wittingly or unwittingly). 

One key advantage of the first option is that the underlying measures behave in a more consistent and predictable fashion. 
In particular, nonlinear effects are sensitive to the parameters of the analysis filter in ways that are difficult to expect, making the second approximation unreliable.
A good example of this is given in section \ref{section:decbias}, where we observed that decentering bias depends on some overlap between the spectrum of the signal envelope and the signal itself.
Decreasing the analysis bandwidth tends to eliminate or attenuate the overlap, making the bias \del{go away }\add{disappear}.
On the other hand, decentering bias masked by correlated amplitude modulation occurs when the analysis filter was sufficiently narrow to isolate individual sidebands arising from the modulation.   
Both of these effects show the sensitivity of the result to the analysis filter.  
In contrast, for stationary cross-spectral estimates, changing the analysis filter bandwidth affects frequency resolution but does not fundamentally change the result.
While broader filters provide a less highly resolved spectral estimate, the decomposition into more finely grained bands remains additive: the picture becomes sharper (and noisier), but it does not otherwise change.

Some authors have cautioned against narrow filters because they artificially force the signal to take on an oscillatory character \cite{yeung2004detection} and create timing ambiguities \cite{vanRullen11}.
A rejoinder to the first point is that techniques which couple filtering with complex demodulation and downsampling matched to bandwidth, \del{such as the DBT, }don't yield an inherently oscillatory signal representation \cite{kovach2015demodulated}.
The equivalence between these techniques and the more standard approach to filtering shows that the oscillations in filtered data are really an artifact of the choice of representation;
the errors associated with narrow filtering come from treating the signal as though it might occupy a wider bandwidth than what has been forced upon it, but these errors may be forestalled through downsampling. With respect to the second point, detailed timing information also remains encoded in the relationship of phase across multiple bands, from which it may be recovered using, for example, an inverse Fourier transform computed across bands.

\subsection{Extensions}
\label{section_extensions}

Amplitude-weighted PLV joins a host of other measures designed to address various shortcomings of PLV and coherence. 
Alternative measures have been proposed to better cope with small sample bias \cite{vinck2010pairwise}, volume conduction \cite{stam2007phase}, as well as decentering bias \cite{van2015phase} and combinations of these problems \cite{vinck2011improved}.
Each of these proposed alternatives has its strengths and weakness, making it suited to some applications but not others, which is certainly also the case for awPLV. 
Most of these measures are meant to address problems with analytic signals considered in isolation.
One important advantage of linearity-preserving measures derived from cross spectra is that the array of tools afforded by spectral analysis remain applicable; these might serve to isolate signal components of interest or to more clearly identify salient properties of the dependence. For example, the time-domain representation of cross-spectral estimators as impulse response functions may reveal useful information about timing that is not reliably conveyed by phase lags considered in isolation. 
This point becomes especially relevant whenever within-band time resolution is  lost to narrowband filtering.
 It is easy to forget that detailed timing information is nevertheless preserved in the relative phase across multiple bands, which may be recovered through the inverse Fourier transform. 
\del{Of course, it is still the case such 2nd order statistics only address linear dependence and share the limitations inherent in linear estimators.}

\section{Conclusion}
\label{section_conclusion}

We have reviewed some pitfalls common to measures of dependence derived from analytic phase and discussed simple approaches to remedying these. 
The main points are: 
\begin{enumerate}
	\item Obtaining analytic phase involves a nonlinear transformation of the signal that may at times create unexpected biases and complicate spectral interpretation (Section \ref{SpectBias}-\ref{section:ampmod}). 
	\item Pure measures of phase exhibit some poor statistical properties in the presence of \del{white }noise, including spectral biases that grow with sample size. 
	By reducing effective sample size, filtering to limit signal bandwidth provides only an apparent but not an actual remedy for this problem (Section \ref{NastyStats}).    
	\item  Measures of linear dependence taken directly from the cross spectrum, such as coherence, preserve spectral specificity and are often statistically better behaved in comparison to pure-phase statistics  (Section \ref{Remedies}).
	\item One such measure, amplitude-weighted phase locking, avoids the ambiguity that is a major drawback of coherence (Section \ref{section_awPLV}).
	\item Phase is often more informative considered across multiple bands jointly than in isolation, especially when drawing inferences related to timing. (Section \ref{sec_linearity}).
\end{enumerate}

\section*{Acknowledgmemts}
The author wishes to thank Phillip E. Gander, Yuki Kikuchi and Christopher Petkov.


\bibliographystyle{IEEEtran}
\bibliography{bibliography}

\begin{appendices}
\section{Effect of Amplitude Modulation}
\label{appendix:AMspect}
\subsection{Effect of Modulation on the Cross Spectrum}

\subsubsection{Broadband Case} If the analysis window, within which the cross spectrum  is to be estimated,  spans both of the positively and negatively translated range of copies of the original signal, then the conjugate product of two identically modulated signals is given by:
\begin{align*}
\begin{split}
     b^2XY^*& = \\
      A_xA_y &e^{i\Delta\phi_{xy}} \left(
       \frac{1}{2}+\mu_b^2 + 2\mu_b|B|\cos\left(\beta\right) +  \frac{1}{2} |B|^2\cos\left(2\beta\right) 
     \right)
     \end{split}
\end{align*}
For simplicity, we consider the case when the envelope of the modulating signal $|B|$ is independent of $A_x$ and $A_y$, so that 
\begin{equation}
\label{EQ_AMsignalXY}
\begin{split}
     \left<b^2XY^*\right>& = \\
       &\left(\frac{1}{2}\sigma_b^2+\mu_b^2\right)S_{xy}
					 +2\mu_b\mu_{|B|}\left<XY^*\cos\left(\beta\right)\right>\\
					 &+ \frac{1}{2}\sigma_b^2\left<XY^*\cos\left(2\beta\right)\right>
\end{split}
\end{equation}
Thus the cross-spectral estimate now includes two 3rd order moments with the modulating phase, such that whenever the phase difference between $X$ and $Y$, $\Delta\phi_{xy}(t)$, assumes a consistent relationship with any of $\pm2\beta(t)$ or $\pm\beta(t)$, it will contribute to the estimate. Correlations between the envelopes $A_x$ and $A_y$ and the modulating phase will also  influence the result.

Such dependence between the signals and their common source of modulation may influence coherence, but the interpretation is clear: it reflects the fact that the cross-spectral measures are weighted according to amplitude. Such measures will therefore tend to reflect the characteristic phase relationship when the product of the signal amplitudes is large.  

On the other hand, if there is no relationship between $B$ and either of the signals or their amplitudes, the outcome is only a rescaling of the original cross-spectrum:
\begin{equation}
\left<b^2XY^*\right> =\left(\frac{1}{2}\left(\sigma_b^2 +\kappa_1+\kappa_2 \right) + \mu_b\right)S_{xy}
\end{equation}
where $\kappa_j = \left<\cos(j\beta)\right>$. In the large-sample limit, the effect of such constant scaling will cancel from coherence (although we will later see that $B$ still plays a role in small sample bias). 

\subsubsection{Sideband Case} If the analysis window only spans one of the shifted spectral bands arising from modulation with $b$, then the outcome of filtering resembles 
\[     bX =   \frac{1}{2}A_x|B| e^{i\Delta\phi_{xy}+i\beta}  
\]
The complex argument of $B$ cancels out in the cross-spectral estimate giving
\begin{equation}
\label{EQ_AMsignalPHnarrow}
\begin{split}
     \left<|B|^2XY^*\right> = \frac{1}{4} \sigma^2_bS_{xy}
  \end{split}
\end{equation}
here assuming independence between $|B|$ and the phase difference $ \Delta\phi_{xy}(t)$.
In calculating coherence, the constant multiplicative terms above will cancel out in the large-sample limit, although as mentioned, modulation may still affect small-sample bias.

\subsubsection{Baseband case} If the analysis window covers only the baseband created by any nonzero constant offset $\mu_b$, the result is unchanged except for being scaled by $\mu^2_b$.

\subsection{Effect of Modulation on Phase Locking}

\subsubsection{Broadband Case} Again, when the analysis \del{band}\add{window} spans the full range of the modulated signal, we have
\begin{equation}
\label{EQ_AMsignalPHnarrow}
\begin{split}
     \Phi_{bX} = \frac{bX}{|bX|}= \frac{b}{|b|}e^{i\phi_x}  =\sgn(b)e^{i\phi_x}
  \end{split}
\end{equation}
If $b$ is strictly positive then modulation clearly has no effect on the  distribution of phases or phase differences between two signals. If the value of $b$ may become negative, phase undergoes a $\pi$ rad flip with every sign change. 
As a result, the phase distribution for each signal will become more symmetric and the mean phase vector will tend towards zero as the probability of $b$ assuming either sign approaches 0.5. 
However, in this case there is still no effect on the resulting phase difference, as 
\[ \Phi_{bX} \Phi^*_{bY}= \left(\sgn(b)\right)^2e^{\Delta\phi_{xy}} = e^{\Delta\phi_{xy}}\quad\mathrm{  a.s.}\] 
In that case, recentering will provide no remedy for bias. 
The $\pm\pi$ jumps occurring at each flip of the sign of B will also contribute to spectral broadening.  

\subsubsection{Sideband Case} As before, when the analysis filter covers only one of the side bands of the modulating signal,  phase is approximated by 
\begin{equation}
\label{EQ_AMsignalPHbroad}
\begin{split}
     \Phi_{BX} = \frac{BX}{|BX|}= e^{i\phi_x+i\beta}
  \end{split}
\end{equation}
In this case, the modulating term enters only the complex argument of each signal, therefore canceling from the phase difference and leaving measures of phase locking unchanged. 
However, the distribution of the phase of input signals will come to reflect $\phi_x(t) + \beta(t)$, and so will be dispersed if $\phi_x$ and $\beta$ vary independently. Mean centering therefore has little chance of suppressing bias in this case.

\subsubsection{Baseband Case}  
If the analysis filter only covers the baseband, the results are unchanged from the unmodulated signal, as the signal within the analysis band is the same as it was before modulation, except for having been scaled by $\mu_b$. Any decentering bias in this band will therefore be corrected by mean subtraction, as before.

\end{appendices}

\end{document}